\documentclass[%
 reprint,
 amsmath,amssymb,
 aps,
floatfix,
]{revtex4-1}

\usepackage{graphicx}
\usepackage{dcolumn}
\usepackage{bm}

\begin{document}

\title{Nitrogen Contamination Causes High Field Q-Slope (HFQS)\\in Buffered Chemical Polished SRF Cavities}
\thanks{This work was supported in part by the U.S. Department of Energy Office of Science under Cooperative Agreement DE-SC0000661; by the U.S. National Science Foundation under Grant PHY-1102511; by the State of Michigan; and by Michigan State University.}%

\author{D. Luo}
\altaffiliation{Luo@frib.msu.edu}
\author{K. Saito}%
\author{S. Shanab}%
  
\affiliation{%
 Facility for Rare Isotope Beams, National Superconducting Cyclotron Laboratory, Michigan
State University, East Lansing, MI 48824, USA
}%

\date{\today}

\begin{abstract}
Buffered Chemical Polishing (BCP) was the most conventional polishing method for superconducting radio frequency (SRF) Niobium (Nb) cavity surface preparation before the discovery of Electropolishing (EP), which is superior to BCP in high gradient performance. The High Field Q-slope (HFQS) is perfectly eliminated by taking the low temperature bake (LTB) post EP, which guarantees high gradient performance in EP'ed cavities. The mechanism of the HFQS is well understood for EP'ed cavities. On the other hand, there is no common consensus on the HFQS with BCP, since even BCP with LTB does not always resolve the HFQS. BCP is much easier to apply and still an important preparation technology for very complicated SRF structures like low beta cavities. Therefore, overcoming the issue of HFQS with BCP is highly beneficial to the SRF community. This paper mines a large number of available data sets on BCP'ed cavity performance with fine grain, large grain, and even single crystal niobium materials under different experimental settings. We found that all existing explanations for HFQS with BCP are inconsistent with some experimental results, and propounded nitrogen contamination as a new model. We checked that nitrogen contamination agrees with all existing data and nicely explains unresolved phenomena. Combining these evidence, we deduce that nitrogen contamination is the cause of HFQS in BCP.

\begin{description}

\item[PACS numbers]
29.20.\textendash c, 74.25.\textendash q, 74.25.Nf

\end{description}
\end{abstract}

\pacs{Valid PACS appear here}
\maketitle


\section{\label{sec:level1}Introduction}

HFQS is a phenomenon where Q$_0$ (unloaded Q) performance of the SRF cavity begins to exponentially drop with increasing gradient around 15 - 25 MV/m due to heating at RF high magnetic field region (equator area) on the SRF surface \cite{Saito-Qslope,Visentin-BCP}, and it finally limits the acceleration gradient to around 30 MV/m. It happens to all cavities that adopt EP or BCP as surface processing method. HFQS seriously degrades the SRF cavity high gradient performance, thus limiting the final energy of the accelerated particles.

\begin{figure}
\centering
\includegraphics*[width=\columnwidth]{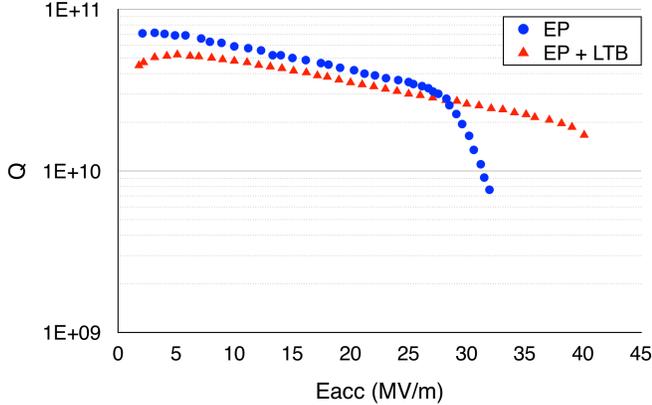}
\caption{An example of High Field Q Slope on EP Nb cavities, before and after baking. HFQS started at 28 MV/m for this cavity. \cite{Saito-Qslope} \label{f:001}}
\end{figure}

\begin{figure}
\centering
\includegraphics*[width=\columnwidth]{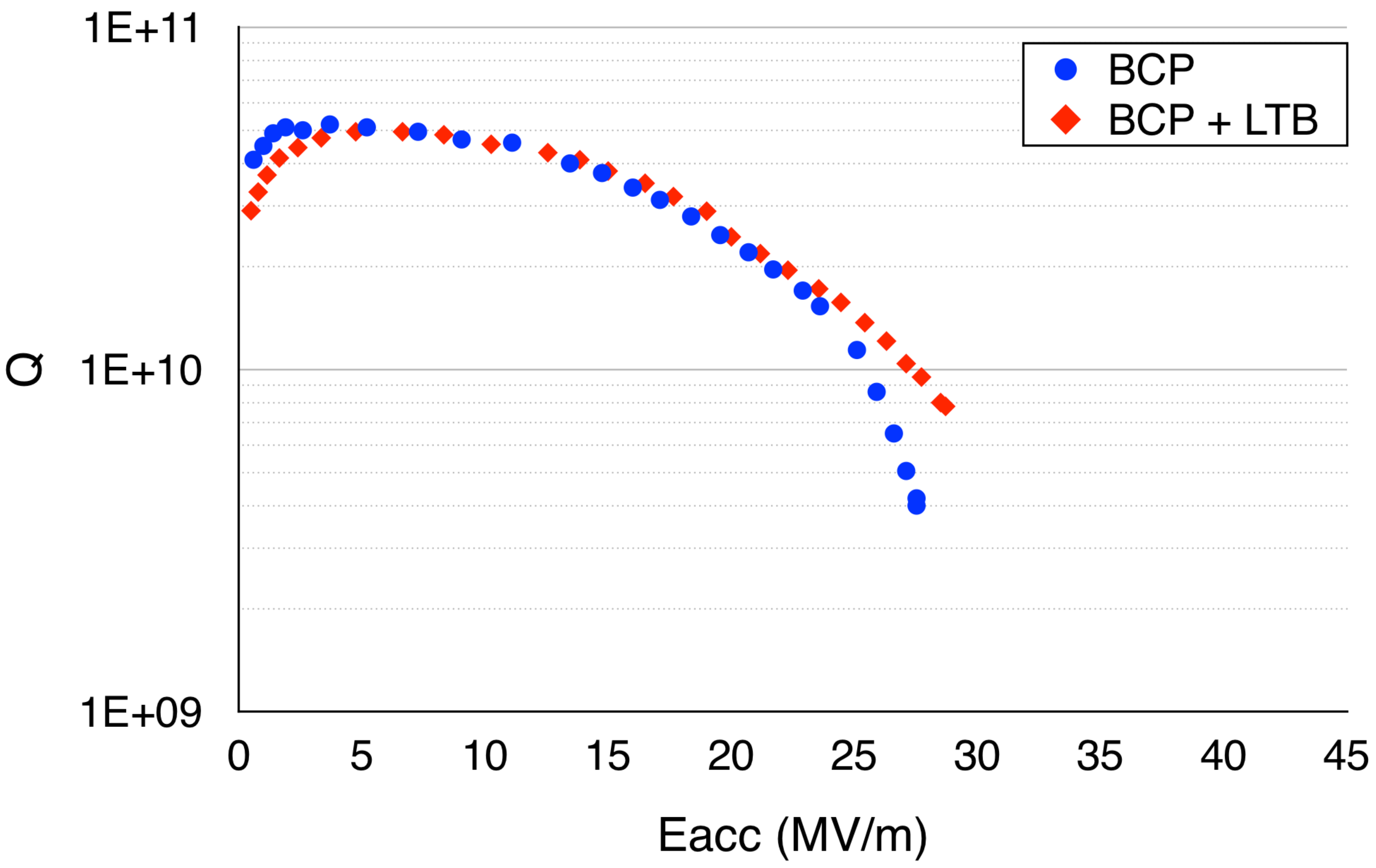}
\caption{High Field Q Slope on BCP Nb cavities, before and after baking. (Courtesy of CEA-Saclay) \cite{Visentin-BCP}\label{f:002}}
\end{figure}

This problem was later neglected after the international linear collider (ILC) chose EP, which was developed in 1985 for KEK TRISTAN 508 MHz 5-cell SRF cavities \cite{Saito-EPDevelepment}, as their baseline preparation method, because EP'ed cavities can completely recover from HFQS via 120 $^oC$ bake for 48 hours \cite{Saito-Qslope}. Fig.~\ref{f:001} shows the Q$_0$ vs E$_{acc}$ (accelerating gradient) of an EP'ed cavity before and after the LTB. However, BCP'ed cavities still suffer from HFQS even after the LTB (see Fig.~\ref{f:002} ).

There is an interesting history behind the LTB for EP'ed cavities. The superiority of EP over BCP with high gradient performance was discovered in 1996 at KEK \cite{saito1998superiority}. They called the high field limitation with BCP'ed cavities as ``European headache'' (nowadays called HFQS) because only European SRF institutes were using BCP and having this problem.

KEK's preparation recipe included the LTB post EP. Its purpose was to improve high vacuum quality to evaporate absorbed water on the SRF surface. They did not yet recognize the important role of LTB for the high gradient cavity performance at that time. KEK transferred their EP method to DESY in 1998, but the first EP'ed cavity at DESY still suffered from HFQS. They revisited the KEK recipe and noticed they were missing the baking process. They applied the bake and finally got high gradient cavity performance \cite{Lilje}. Since then very intensive study started with HFQS in many SRF institutes in the world.

A lot of models for the HFQS exists: thermal feedback \cite{Saito-thermal}, Oxygen contamination \cite{Saito-O,Ciovati,Visentin-BCP,Safa}, field enhancement at grain boundary steps (or macro surface roughness, in this paper we are mostly talking about this roughness) \cite{Knobloch}, field enhancement induced by surface roughness in the grain (or micro surface roughness) \cite{Xu}, interface tunneling exchange \cite{Halbritter}, Hydrogen trapping \cite{Romanenk}, and flux trapping \cite{Ben-fluxtrap}. A good summary is shown in the reference \cite{Hasan2chp5}. Among these models, the Oxygen contamination model can sufficiently explain HFQS in the LTB effect with EP'ed cavities \cite{Visentin-BCP}; however, it cannot explain why LTB fails to totally remove the HFQS for BCP'ed cavities. A fully consistent model of HFQS with BCP'ed cavities does not exist even now.

HFQS study with BCP'ed cavities in early 2000's was for fine grain niobium cavities. Nowadays, many BCP'ed cavity performance information is available for large grain and even single crystal niobium cavities. This paper will mine both the past data with fine grain cavity, as well as studies on large grain and single crystal niobium cavities, in order to develop and prove a new model which can consistently explain the BCP'ed cavity HFQS.

\begin{figure}
\centering
\includegraphics*[width=\columnwidth]{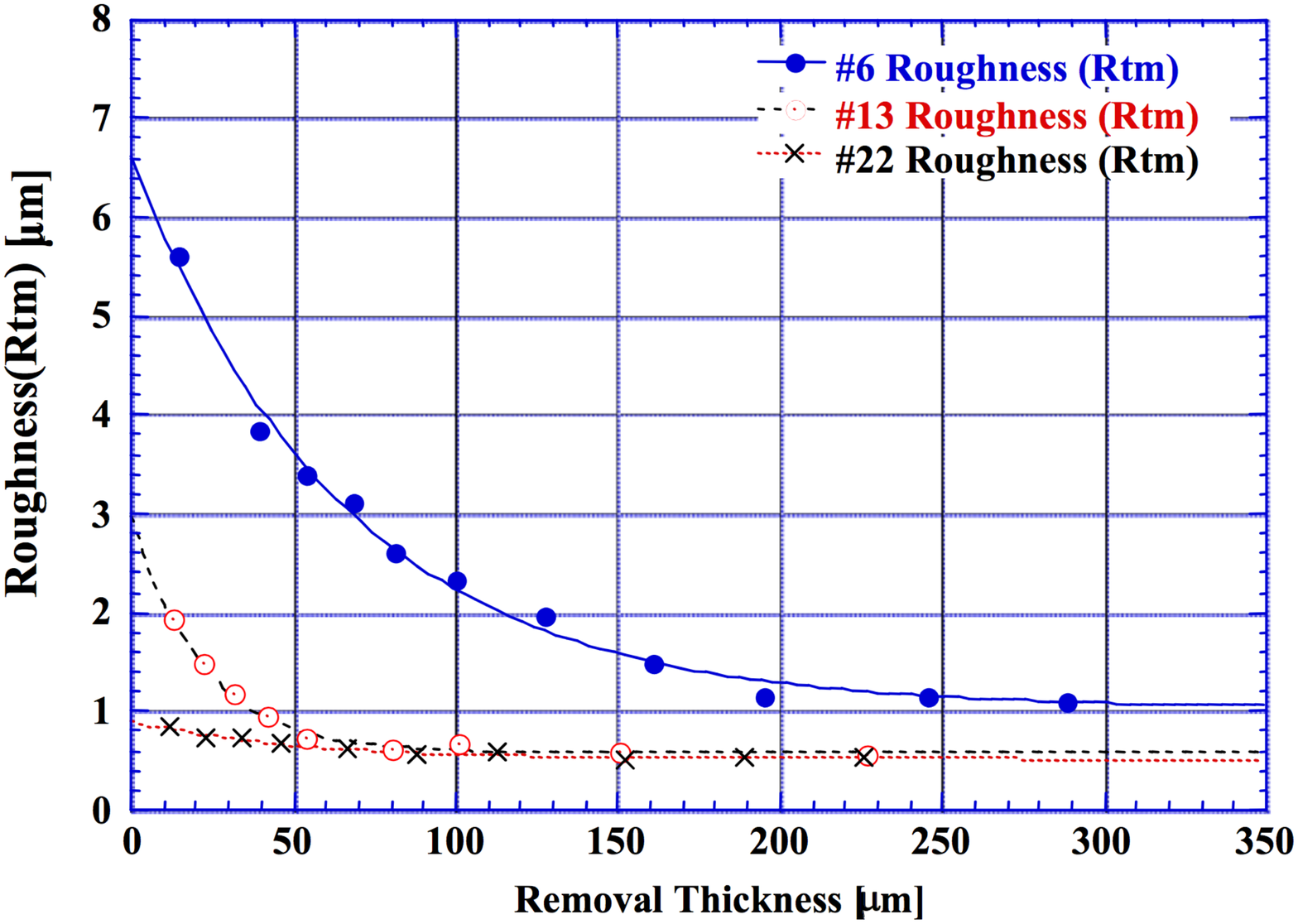}
\caption{ Variation of the surface roughness by EP for fine grain Nb. \cite{Saito-Roughness}\label{f:003}}
\end{figure}

\begin{figure}
\centering
\includegraphics*[width=\columnwidth]{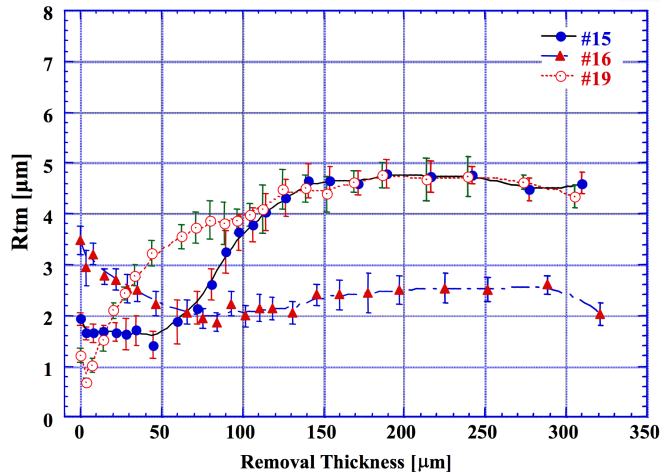}
\caption{Variation of the surface roughness by BCP for fine grain Nb. \cite{Saito-Roughness}\label{f:004}}
\end{figure}

\section{Data Mining}

\begin{figure}
\centering
\includegraphics*[width=\columnwidth]{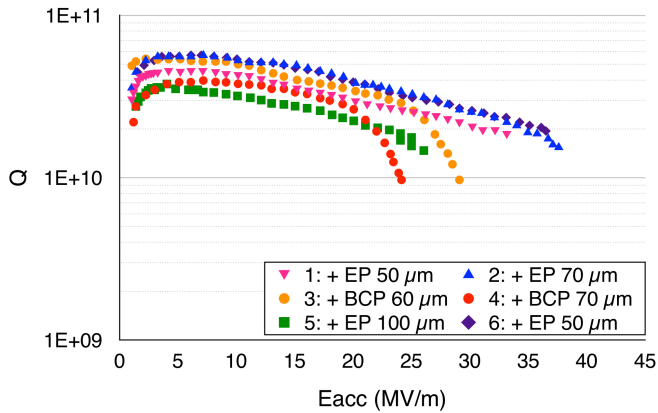}
\caption{Q vs E results for 6 different polishing treatments, all cavities are tested after HPR and baking in KEK. \cite{Saito-EPBCP}\label{f:005}}
\end{figure}

As mentioned above, in the early 2000's HFQS studies with BCP'ed cavities were mostly for fine grain niobium cavities. Discussion of the results centered around the topography of niobium surface treated by BCP, as a natural consequence the effects of surface roughness (Rtm, height difference between 3rd highest peak to 3rd lowest minimum), grain boundaries, defect and dislocation were the main concerns. Fig.~\ref{f:003} and \ref{f:004} show the polishing characteristics of EP and BCP respectively \cite{Saito-Roughness}. EP makes the surface smoother exponentially with increased material removal, while BCP makes the surface smooth in the early stage but gradually rougher with increased material removal due to the preferential etching at grain boundary areas, and the roughness saturates around 2 - 5 $\mu$m depending on grain size of the niobium material. The experiments in early 2000's established the following facts:
\begin{enumerate}
  \item BCP'ed cavity HFQS is not sufficiently improved by the LTB, the Q-drop is somehow mitigated but the gradient limit sees little improvement (Fig.~\ref{f:002}).
  \item Sometimes the Q-drop is eliminated by the LTB but the field limitation still does not change, unlike EP.
  \item BCP after EP lowers the onset field of the HFQS. Successive BCP makes the HFQS (Fig.~\ref{f:005}) worse.
  \item The degraded HFQS is perfectly recovered by EP with a rather heavy material removal (150 $\mu$m, see Fig.~\ref{f:005}). The degradation of BCP HFQS might be explained by the RF field enhancement on the rougher surface.
\end{enumerate}

\begin{figure}
\centering
\includegraphics*[width=\columnwidth]{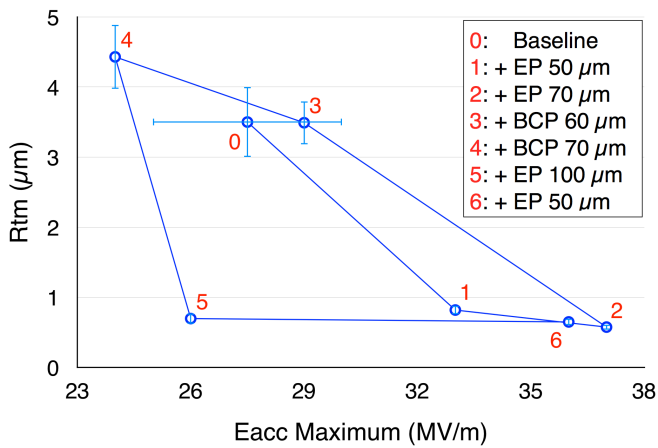}
\caption{Roughness vs E$_{acc, max}$ results for 6 steps of the polishing treatment (each with HPR and baking) in KEK multiple EP and BCP experiment. Each result is marked with the respective number in Fig.~\ref{f:005}. The roughness information is from Table I. \cite{Saito-EPBCP}\label{f:006}}
\end{figure} 

These experimental facts seem to support the field enhanced model on the rough surface \cite{Xu} or grain boundary steps \cite{Knobloch}. However, more detailed analysis does not support these models for BCP HFQS as shown below.

\subsection{\label{sec:level2}Reanalysis of multiple BCP post EP with fine grain cavity}

The cavity inner surface roughness in the experiments of Fig.~\ref{f:005} can be estimated from the surface polishing characteristics in Fig.~\ref{f:003} and \ref{f:004}. Table I shows the results. The calculation assumed initial surface roughness is 3.5 $\mu$m. Fig.~\ref{f:006} also shows the surface roughness vs. the limiting field in the experiment. It has a hysteresis: the EP 100 $\mu$m at step 5 should improve the roughness to less than 1 $\mu$m for most assumption of the initial roughness, which is enough to reach a high gradient $>$ 30 MV/m but in reality the gradient is still limited $<$ 30 MV/m. 

It is very clear that from step 4 to 5, the roughness improves a lot by EP, while the maximum field does not; from step 5 to 6, the roughness doesn't change much while the maximum gradient increases from 26 MV/m to about 36 MV/m. This fact suggests there is something in addition to surface roughness that is causing BCP'ed cavity HQFS.

Getting a little ahead, the improvement from step 5 to step 6 can be explained if the cause of HFQS is nitrogen contamination by the nitric acid in the BCP acid, as will be shown later in detail. The contamination was accumulating (especially through grain boundaries) during multiple BCP, and 100 $\mu$m EP in step 5 is not enough to remove them all.

\begin{table}[b]
\caption{\label{tab:table2}
Estimation of roughness during KEK multiple EP and BCP process.}
\begin{ruledtabular}
\begin{tabular}{lcc}
 Process  &  Rtm ($\mu$m) & Uncertainty ($\mu$m) \\
 \colrule
Baseline        & 3.50   & 0.50 \\
+ EP 50 $\mu$m + baking  & 0.82  & 0.06 \\
+ EP 70 $\mu$m + baking  & 0.57  & 0.04 \\
+ BCP 60 $\mu$m + baking & 3.49  & 0.31 \\
+ BCP 70 $\mu$m + baking & 4.43  & 0.46 \\
+ EP 100 $\mu$m + baking & 0.70  & 0.07 \\
+ EP 50 $\mu$m + baking   & 0.65 & 0.05

\end{tabular}
\end{ruledtabular}
\end{table}

\subsection{HFQS of BCP'ed large grain cavity}

BCP has preferential etching on grain boundary steps. During the BCP process, roughness in the grain is getting lower while the gain boundary step difference is getting larger. The grain size of fine grain cavities is $\sim$ 50 $\mu$m \cite{Hasan1Chapt6finegrain}. Surface roughness is usually measured by a needle scanning a 0.8 mm width, so the measured surface roughness with fine grain Nb materials combines topography in the grain and on grain boundary step.

To date large grain sheets are available by directly slicing large grain Nb ingot. The grain size is on several cm scale. The measured BCP finishing surface roughness is $\sim$ 0.1 $\mu$m, it is very smooth (lower than EP $\sim$  0.5 $\mu$m) and even mirror-like. Measurements of large grain surface roughness count the roughness in the grain only, because the grain size is beyond measurement range. 
Considering the BCP property, the large grain smoothness is mostly due to the fact that there is much less grain boundaries, while the roughness in the grain doesn't change much. 

\begin{figure}
\centering
\includegraphics*[width=\columnwidth]{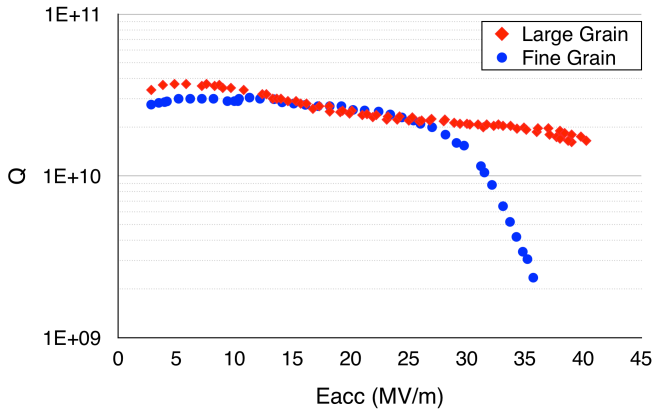}
\caption{ Comparison of quality factor behaviours at high gradients between the Large Grain (LG) and Fine Grain (FG) cavities by BCP with baking. \cite{IHEP}\label{f:007}}
\end{figure}

Fig.~\ref{f:007} shows a comparison of the cavity performance between BCP fine grain and large grain cavities \cite{IHEP}. The onset of BCP HFQS is pushed up to $>$ 40 MV/m, higher than that of fine grain cavities. This experiment shows less grain boundary steps give much better performance, or more grain boundary steps give worse performance, thus indicating the grain boundary is responsible for the BCP HQFS. The surface roughness in the grain (micro roughness) is ruled out as a cause of BCP HFQS by this experimental result.

\subsection{BCP after EP with large grain cavity}

DESY investigated the BCP impact post EP for large grain cavity \cite{Singer-2}. Fig.~ 8 shows the result. BCP (50 $\mu$m removal) degrades remarkably the high gradient performance, and is similar to the BCP fine grain cavities.

\begin{figure}
\centering
\includegraphics*[width=\columnwidth]{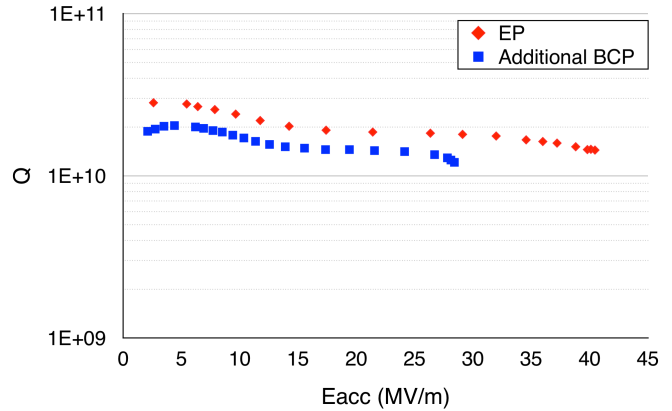}
\caption{ Q vs E of DESY Large Grain TESLA shape cavity treated with EP and an additional 50 $\mu$m of BCP. Both results are after baking. (Courtesy of DESY $\&$ JLab) \cite{Singer-2}  \label{f:008}}
\end{figure} 

\subsection{Multiple BCP with large grain cavity}

\begin{figure}
\centering
\includegraphics*[width=\columnwidth]{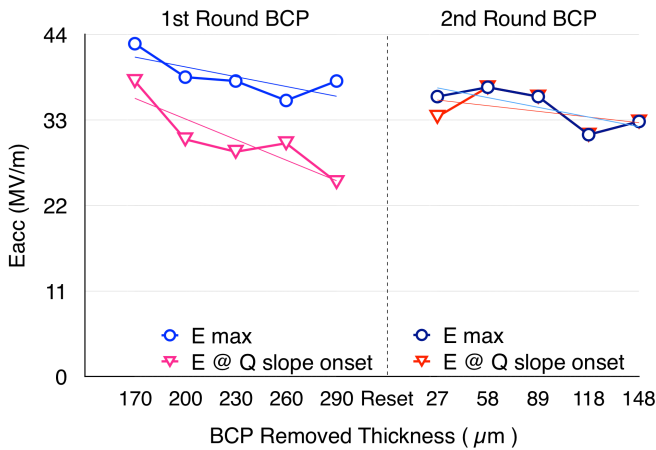}
\caption{KEK $E_{acc}$ vs Increasing BCP amount results, first round and second round. All results were taken after LTB. \cite{Furuta-CBP} \label{f:009}}
\end{figure}

KEK investigated the impact of successive BCP on large grain cavity \cite{Furuta-CBP}. A 1.3 GHz Ichiro shape cavity was mechanically polished inside ($\sim$ 80$\mu$m by the centrifugal barrel polishing (CBP)). Note that Ichiro cavity shape has a smaller B$_p$/E$_{acc}$ ratio (3.56 mT/[MV/m]) compared to the ILC shape (B$_p$/E$_{acc}$ = 4.26 mT/[MV/m]). After 750 $^oC$ for 3 hrs hydrogen degas annealing, a bulk etching of 160 $\mu$m took place using vertical BCP (V-BCP, conventional BCP in most institutes).  The tight loop test took place in total 5 times removing 30 $\mu$m each time by V-BCP (first cycle). In these test the LTB is applied post V-BCP for all measurement. The result is shown in Fig.~\ref{f:009}. The decreasing of E$_{acc}$ onset of HFQS is observed with increased V-BCP material removal. In the second cycle, the surface topography was reset by CBP, then the tight loop test was repeated by H-BCP. The original experimental goal was to validate that the horizontal BCP (H-BCP) performance is better than normal V-BCP due to more uniform material removal. Again the degradation of onset HFQS happened for each extra 30 $\mu$m BCP, the average of the E$_{acc}$ onset of HFQS (red line) increased compare to the 1st cycle (pink line), but E$_{acc,max}$ average (dark blue line) is lower than the 1st cycle (blue line). The conclusion from these results is that resetting surface roughness doesn't work to improve the HFQS for E$_{acc,max}$, which suggests lowering the (macro) surface roughness doesn't sufficiently increase the BCP maximum gradient.

Temperature map result for a large grain cavity also shows that there is no preferential heating on the grain boundary area \cite{Grainboundary-heating}, where the high roughness region is mainly located.

\subsection{BCP with single crystal cavity}

\begin{figure}
\centering
\includegraphics*[width=\columnwidth]{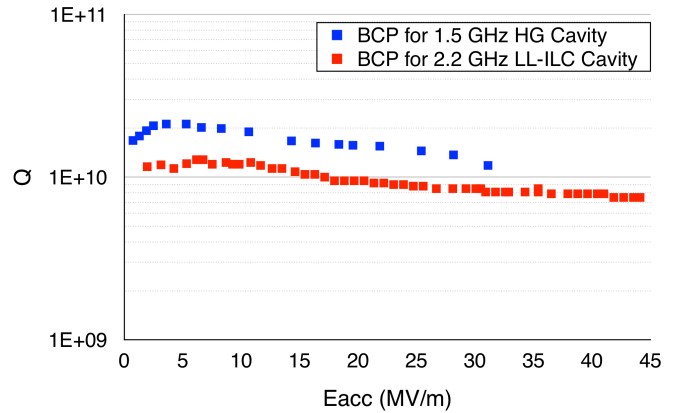}
\caption{Q vs E of the 1.5 GHz High Gradient (HG, blue curve, $B_p/E_{acc}$ is 4.47 mT/(MV/m)) shape and 2.2 GHz Low Loss-ILC (LL-ILC, red curve, $B_p/E_{acc}$ is 3.56 mT/(MV/m)) shape Single Crystal (SC) Cavities treated by BCP. Both results are after baking. (Courtesy of JLab) \cite{Kneisel-single,Kneisel-Large} \label{f:010}}
\end{figure}

Single crystal cavity can completely rule out concern of the topographic issue with BCP HFQS, which has no grain boundary and can attain 0.1 $\mu$m scale very smooth surface even by BCP. To date single crystal niobium sheets are available via directly slicing the ingot with very large grain at 15 - 30 cm scale. Cavities are assembled by electron beam welding (EBW), however it does not produce any grains if the crystal orientation is arranged with the two half cells \cite{Saito-singlecry}. So we can fabricate the entire interior cavity surface with a single crystal. P. Kneisel fabricated a 2.2 GHz single crystal Low Loss-ILC shape cavity cutting out single crystal sheets from the large gain ingot (CBMM material), treated it with BCP plus baking and investigated the performance \cite{Kneisel-single}. The result is shown in Fig.~\ref{f:010} red curve. We found that combining BCP and LTB for single crystal cavity, the HFQS disappeared like normal EP cavities, while the maximum gradient is still limited. The LTB eliminates the HFQS but field gradient still limited $\sim$ 45 MV/m (160 mT) and did not reach the fundamental field limit ($\sim$ 180 mT) unlike EP'ed cavities. This provides direct evidence to rule out the topography as a major cause of BCP HFQS.

A nearly single crystal 1.5 GHz High Gradient shape cavity shows similar conclusion. This cavity was fabricated with a Nb ingot that has a $\sim$ 20 cm diameter grain in the center and several small grains $\sim$ 1 cm. It is treated with BCP plus LTB , with very smooth surface on most part of the cavity. The field limit is 143 mT as shown in Fig.~\ref{f:010} blue curve.

These evidence shows that BCP cannot reach the fundamental field limit like EP and suggests there is still something else limiting the gradient in BCP.

\begin{figure}
\centering
\includegraphics*[width=\columnwidth]{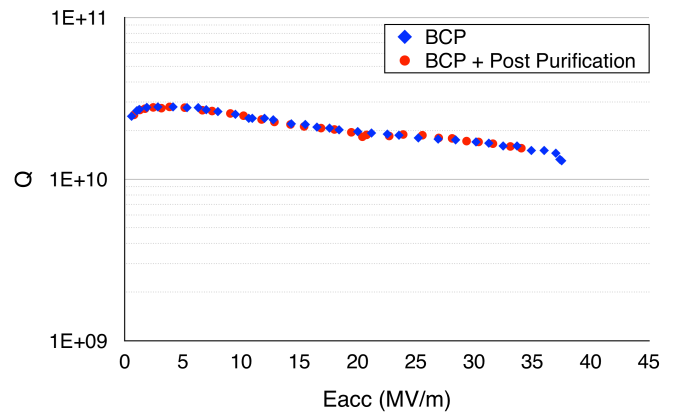}
\caption{Q vs E DESY Single Crystal TESLA cavities treated with BCP: before post-purification \cite{Singer-1} in blue curve; after post-purification in red curve. Both results are after baking. (Courtesy of DESY $\&$ JLab) \cite{Myneni} \label{f:011}}
\end{figure}

The same experiment took place for 1.3 GHz ILC type single cell cavity fabricated at DESY \cite{Singer-1}. The single crystal sheets were produced by Heraeus. Fig.~\ref{f:011} blue curve shows the result of BCP (112 $\mu$m and 120 $^oC$ bake for 6 hours). Q curve is very flat up to the gradient limit. The gradient is limited by 37.5 MV/m ($\sim$ 160 mT) which is close but still below the fundamental limit 180 mT. 

One important fact is that this performance is worse than the DESY large grain TESLA shape cavity treated with EP in Fig.~\ref{f:008}. This goes against the common understanding: single crystal cavity, even treated with BCP, has very smooth surface and better thermal conductivity, so it should have better performance than large grain cavity. This indicates the surface roughness itself cannot explain all the performance degradation.

Post-purification was also tested for this cavity. The result is shown in Fig.~\ref{f:011} red curve. After the post-purification the cavity was BCP'ed and took 120 $^oC$ bake for 12 hrs. This process does not improve the field limit. 
The performance failed to improve even with reduced stress or improved dislocation on the SRF surface by post-purification. This also suggests BCP still contains other issues.

\section{Nitrogen Contamination}

By the above data mining, we can conclude that the topographic character of the SRF surface produced by the BCP polishing feature isn't the only cause of the HFQS with BCP'ed cavities. Therefore, we need to look for other causes. 

This paper focuses on finding why BCP can't get enough improvement as EP from LTB, so we may only focus on the differences between those two methods. As shown in Table II, EP and BCP difference: roughness (discussed before and rule out) and potential contamination (H, O, C, and S for EP; H, O, C, N and P for BCP). F from HF is proved not harmful for cavity performance, and P in phosphoric acid has never been reported to react with Nb. Then the only contamination difference is N and S. S problems were discussed in several papers and solved, therefore N becomes the only potential contamination that causes worse performance in BCP.

\begin{table}[b]%
\caption{\label{tab:table1}%
Main differences between EP \& BCP
}
\begin{ruledtabular}
\begin{tabular}{ccc}
\textrm{ }&
\textrm{EP}&
\textrm{BCP}\\
\colrule
R$_z$ ($\mu$m) & 0.5 & 2 \\
\colrule
Grain Boundary  \\
Prefer Etching & No & Yes\\
\colrule
Contamination & H, F, O, C, S & H, F, O, C, N, P \\
\end{tabular}
\end{ruledtabular}
\end{table}

\begin{figure}
\centering
\includegraphics*[width=\columnwidth]{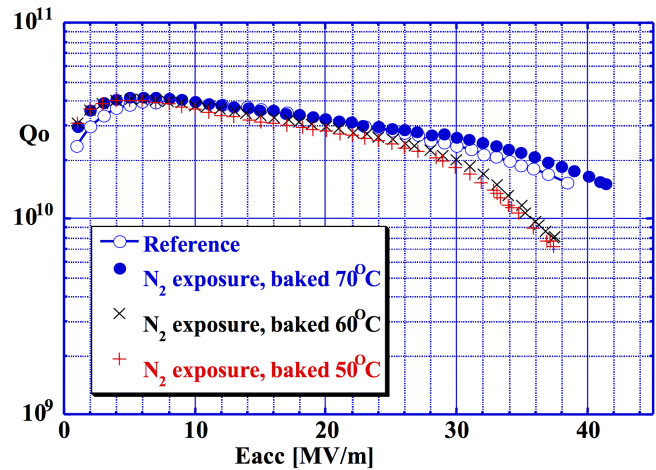}
\caption{KEK cavity performance degradation with nitrogen exposure \cite{Saito-N} \label{f:012}}
\end{figure}

Our first attention to the nitrogen contamination was the gas exposure test on the fresh SRF surface just after vertical test, without vacuum break before the gas exposure \cite{Saito-N}. We found that usually argon gas exposure has no performance degradation but pure nitrogen gas exposure produces a remarkable Q degradation as shown in Fig.~\ref{f:012}. In this case, the low temperature (70 $^oC$) bake even lower than 120 $^oC$ seems an effective cure. We also notice that a similar experiment shows the opposite result: exposing Nb cavity to nitrogen gas for three days has no degradation observed in their case \cite{N-exposure}.

\begin{figure}
\centering
\includegraphics*[width=\columnwidth]{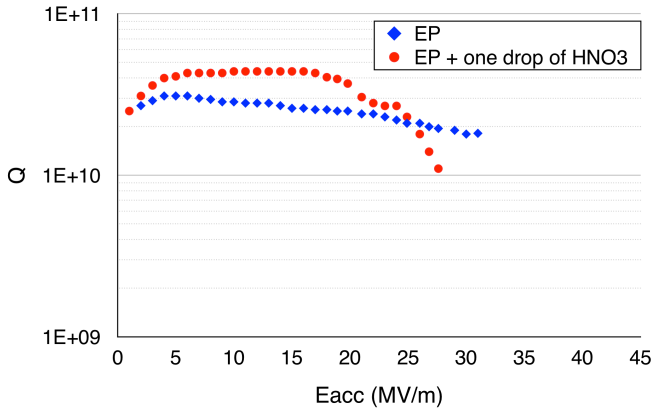}
\caption{ Q vs E for KEK fine grain Type A cavity, blue cross curve is an example for normal EP cavity performance \cite{Saito-EPbetter},  red circle curve is the cavity dealer with EP plus one drop of nitric acid \cite{Higuchi-p}. Both cavities was applied baking after EP. \label{f:013}}
\end{figure}

Our hypothesis that nitrogen contamination produces the HFQS is first corroborated by the experimental results summarized in Fig.~\ref{f:013}. This experiment was originally done for development of hydrogen-free EP \cite{Higuchi}. Adding nitric acid of (61\% w/w) 1500 ppm into the EP acid (48\% HF: 93\% H$_2$SO$_4$ = 1: 10 V/V) was very effective to prevent hydrogen doping during EP. However, unlike the normal EP case, the HFQS still exists even after baking. The only difference between these two processes is the small amount of the nitric acid, so that we think the N contamination gives this degrading of the cavity field limit.

The second decisive evidence is in the nitrogen doping study. FNAL has successfully developed the nitrogen doping method to increase Q of SRF Nb cavities \cite{Grassellino}. They put nitrogen gas into the vacuum furnace at 800 - 900 $^oC$ vacuum annealing to contaminate the surface. They proved that only the interstitial nitrogen in the niobium contributes to enhance the Q because it shortens the mean free path. During the doping process, niobium-nitride metal phase (Nb$_x$N$_y$) is also generated on the top surface. This has to be removed by several microns of electropolishing, otherwise remarkable field limit happens. Later, low temperature ($\sim$ 120$^oC$) nitrogen infusion was developed \cite{grassellino2017unprecedented}. The cavities treated with this method have much higher maximum gradient, in the mean time niobium nitride doesn't exist, this further indicates nitrogen contamination will lower the gradient.

Thirdly, KEK experiment with stepwise BCP and reset (Fig.~\ref{f:009}) indicates that, if the HFQS is attributed to N contamination, the CBP before second experiment cycle only resets the roughness, but does not remove all the contamination, or BCP after CBP already produced deeper N contamination, and this explains the degradation of the E max.

Finally, the KEK experiment with multiple EP followed by BCP and then more EP (Fig.~\ref{f:005}) also agree with this assumption. After the step 5, 100 $\mu$m EP removal for Nb is sufficient for recover the roughness to less than 1 $\mu$m, however, it is not enough to remove the contamination in the deep site, so the maximum gradient is limited to only $\sim$ 26 MV/m. 

While after the step 6, even though the roughness does not change much, it removes 50 $\mu$m more Nb which contains most N contamination. Therefore, the E$_{acc,max}$ increased a lot, and is close to the original result $\sim$ 37 MV/m.

\section{GENERATION OF NITROGEN
CONTAMINATION AND OBSERVATION}

\begin{figure}
\centering
\includegraphics*[width=\columnwidth]{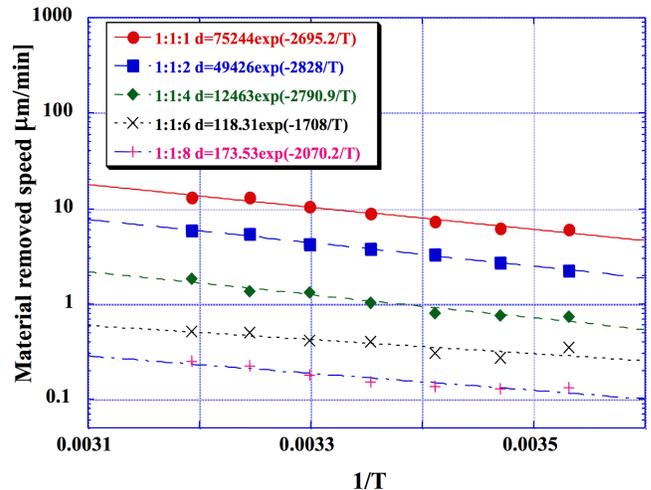}
\caption{Reaction speed vs 1/T for different BCP ratio. \cite{Saito-BCPEnergy} \label{f:014}}
\end{figure}

During nitrogen doping, Nb$_x$N$_y$ starts to generate at temperature $\sim$ 400 $^oC$ (\cite{Nitride}, is 673 K, corresponds to 0.058 eV). Nb$_x$N$_y$, which has a very low thermal conductivity (1/10 of Nb at low temperature), is harmful for cavity performance, and could be the reason for low E$_{acc}$ and HFQS. While the low temperature (120 $^oC$) infusion doesn't form Nb$_x$N$_y$ so this can produce high gradient and high Q cavity performance \cite{Gra-Nitrogeninfusion}.
We can then assume that if Nb$_x$N$_y$ can be generated on Nb surface on BCP, it can cause HFQS. The reaction energy of BCP is estimated by measuring the temperature dependence of the polishing rate \cite{Saito-BCPEnergy}. The result is shown in Fig.~\ref{f:014}. The results are well fit by Arrhenius equation:
\begin{equation}
    d(T)=d_0exp(-\frac{Q}{k_BT})
\end{equation}

The constant value of 2695.2 for the BCP acid composition HF: HNO$_3$: H$_3$PO$_4$ = 1: 1: 1 (V/V) corresponds to 0.232 eV ($\sim$ 2419 $^oC$). Nb$_x$N$_y$ can form (but with very low probability reaction) by the nitric acid reacting with niobium in such an energy. Actually JLAB has observed nitrogen by SIMS on BCP'ed niobium sample \cite{Tuggle}. Nitrogen element mostly stays on the surface 0.05 $\mu$m. On the other hand, FNAL showed that interstitial nitrogen and Nb$_x$N$_y$ phase lie at different depths on niobium top surface. Nb$_x$N$_y$ mostly resides on the surface within ~2 $\mu$m \cite{Romanenko}. These indicate the niobium contamination on Nb top surface by BCP is in Nb$_x$N$_y$ phase.

\section{Conclusion}

HFQS with BCP'ed cavities is not perfectly removed by the LTB unlike EP'ed cavities.  There is no common consensus on the mechanism of the HFQS in BCP'ed cavities to date. We mined past data on fine grain, large grain and single crystal cavities to show inconsistencies in previous explanations, and developed a new model for the root cause of the BCP HFQS.  This model says that the nitrogen contamination especially niobiumÐnitride phase (Nb$_x$N$_y$) is the root cause of the HFQS BCP'ed cavities. The niobium-nitride phase is generated by reaction between niobium and the nitrogen from the decomposed nitric acid in the BCP reaction. Nitrogen in niobium-nitride phase does not move by the LTB, thus the HFQS is not recovered by the LTB. Not only does the nitrogen contamination model agree with all experimental data sets, it also offers coherent explanations for previously unresolved phenomena in circumstances including: (1) applying multiple BCP post EP (Fig.~\ref{f:005} and Fig.~\ref{f:006}); (2) adding HNO$_3$ to EP (Fig.~\ref{f:013}); and (3) maximum field of BCP single crystal cavities ( Fig.~\ref{f:010} and Fig.~\ref{f:011}) lower than EP large grain cavity ( Fig.~\ref{f:008} ). These results provide very strong evidence to prove that our model is correct.

\section{acknowledgement}

The authors are grateful to fellow colleagues in the SRF community, especially the KEK SRF group, whose numerous experiments on BCP and EP made our investigation possible, with special thanks to Bernard Visentin (CEA-Saclay), Peter Kneisel (JLAB) and Waldemar Singer (DESY) for allowing the authors to use their figures. We would also like to express our gratitude to Yoshishige Yamazaki (FRIB) and Steven Lund (FRIB) for many fruitful discussions and kind suggestions.

Work supported by the U.S. Department of Energy Office of Science under Cooperative Agreement DE-SC0000661; by the U.S. National Science Foundation under Grant PHY-1102511; by the State of Michigan; and by Michigan State University.

\end{document}